\documentclass[11pt,draftcls,onecolumn]{IEEEtran}
\usepackage{graphicx}
\usepackage{bm}
\usepackage{url}
\usepackage{amsfonts,amssymb,amsmath} 
\usepackage{mystyle}
\usepackage{flushend}
\usepackage{setspace}
\addtocounter{MaxMatrixCols}{50}
\allowdisplaybreaks

\begin{document}

\title{Rate-Compatible Short-Length Protograph LDPC Codes}

\author{ Thuy Van Nguyen, {\em Student Member, IEEE} and Aria
    Nosratinia, {\em Fellow, IEEE}
\thanks{The authors are with the
  Department of Electrical Engineering, The University of
  Texas at Dallas, Richardson, TX 75083-0688 USA Email: nvanthuy@utdallas.edu, aria@utdallas.edu}}

\maketitle

\begin{abstract} 
This paper produces a rate-compatible protograph LDPC code at $1$k
information blocklength with superior performance in both waterfall
and error floor regions. The design of such codes has proved
difficult in the past because the constraints imposed by structured design
(protographs), rate-compatibility, as well as small block length, are
not easily satisfied together. For example, as the block length
decreases, the predominance of decoding threshold as the main
parameter in coding design is reduced, thus complicating the search
for good codes. Our rate-compatible protograph codes have rates
ranging from $1/3$ to $4/5$ and show no error floor down to $10^{-6}$
FER.
\end{abstract}
\section{Introduction}

A rate-compatible family of codes consists of a nested structure that
allows operation in a range of rates using a common encoder/decoder
infrastructure~\cite{Hagenauger_88,ha:IT04,pishro:ITW04,Tian:EURASIP05,Amir:TComL04,ha:IT06,Hyo:Twireless07}. This
nested structure is desirable for many applications, e.g., hybrid
automatic repeat request where a transmitter sends
incremental parity bits to help a receiver to decode a packet received
in the previous time slot.

Rate-compatible codes are usually designed via puncturing, i.e., by
starting from a low-rate code and successively discarding (puncturing)
coded
bits~\cite{ha:IT04,pishro:ITW04,Tian:EURASIP05,ha:IT06,Hyo:Twireless07,El-Khamy2009}. Despite
its simplicity, puncturing is not free of problems. In practice,
puncturing has often produced high-rate LDPC codes whose performance is
inferior to stand-alone codes of similar rate. Also, the low-rate code
and puncturing patterns are often designed separately, which is
suboptimal~\cite{jacobsen:VTC07}.  In this letter we use an approach
involving code extension that starts with a high-rate code (a
so-called daughter code) and obtains low rate codes by extending the
parity matrix of the high-rate
code~\cite{jacobsen:VTC07,Li:02,dolinar:ISIT05, Amir:TComL04}. Most
existing extension-based rate-compatible LDPC
codes~\cite{jacobsen:VTC07,Li:02, Amir:TComL04} are designed for
random-like (un-structure) codes that do not promote low encoding
complexity. We focus on protograph-based LDPC codes, a structured LDPC
code that is built from a small graph known as a
protograph~\cite{Thorpe2003}. This structured code has many advantages,
including near-capacity iterative decoding thresholds, low encoding
complexity allowed by circulant permutations, as well as fast
decoding~\cite{Thorpe2004}. One of the successful examples of
protographs is the AR4JA codes of~\cite{Divsalar2009} which will play
a role in this paper as well. AR4JA codes are nested, but they are not
rate-compatible due to the lack of a constant data block length.

This paper concentrates on short/moderate blocklength codes,
concentrating on providing good waterfall and error floor performance
in the low blocklength regime, which has been a distinct challenge.
The codes produced in this paper are, broadly speaking, related to the
long-blocklength codes designed in~\cite{Thuy:TCOM10}. Specifically,
certain ideas involving the reduction of protograph search space are
borrowed from~\cite{Thuy:TCOM10}. However, the technique
of~\cite{Thuy:TCOM10} was not able to produce good codes at low
blocklengths due to its reliance on the decoding threshold and
minimum-distance growth properties of the protograph.  These two
parameters are good measures for codes with long blocklengths, but for
short/medium blocklengths, they do not possess the same significance.
In this paper we use a mixed cost function involving both the above
mentioned parameters as well as the actual code performance at a given
error probability. Starting from the rate-$4/5$ AR4JA code
of~\cite{Divsalar2009} and building on it via extensions, we are able
to produce a family of rate-compatible codes with waterfall and error
floor performances superior to existing structured rate-compatible
LDPC codes with $1$k information blocklength. At frame error rate
(FER) of $10^{-5}$ our codes exhibit a gap of only $1.7$ dB from the
capacity limits. Simulations down to FER of $10^{-6}$ reveal no error
floors.


\section{Design Method}
\label{sec:design}

A protograph contains a small number of variable and check nodes that
are interconnected via edges. Parallel edges are allowed in a
protograph. A protograph {\em code} is an LDPC code built from the
protograph via lifting, which is the process of copying the protograph
repeatedly and permuting the edges corresponding the same node
type across different copies to interconnect them~\cite{Divsalar2009}.

\begin{figure}
\centering
\includegraphics[width=2.5in]{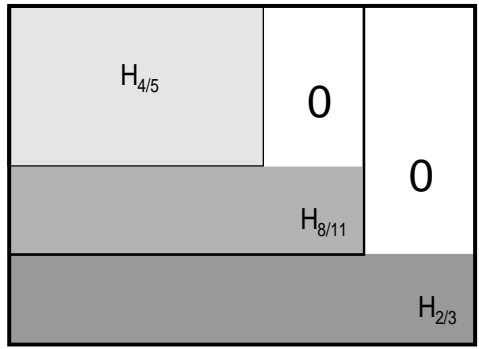}
\caption{Designing rate-compatible codes by extension of the parity check matrix}
\label{fig:rc-structure}
\end{figure}

A rate-compatible set of parity check matrices is shown in
Figure~\ref{fig:rc-structure}. This work starts with a high-rate code
known as the daughter code. A lower-rate code is obtained by adding
the same number of columns (variable nodes) and rows (check nodes)
into the daughter code parity check matrix. The new variable nodes are
connected only to the new check nodes, so that high-rate codewords are
nested in the low-rate codewords.

This process needs a good daughter code to start with. Experiments show
that for short/medium blocklengths, a high-rate protograph should have
a small graph size, because this allows a bigger permutation matrix so
that it is possible to avoid short cycles in its derived LDPC graph
via lifting process.  We found that the rate-$4/5$ AR4JA
code~\cite{Divsalar2009} is an excellent candidate to build
on.\footnote{The highest-rate member of the AR4JA code has
  rate-$9/10$, but that protograph has 10 additional nodes, which is
  contrary to the desirable property of starting with a small
  protograph for our daughter code.} The AR4JA code has low iterative
decoding threshold, linear minimum distance growth property as well as
decent finite-length performance~\cite{Divsalar2009}. We note that
although AR4JA codes are nested, they are not rate-compatible since
the information blocklengths of their nested components are not
identical. Therefore the AR4JA family as a whole is not a candidate
solution to the problem posed in this paper.

\begin{figure}
\centering
\includegraphics[width=3in]{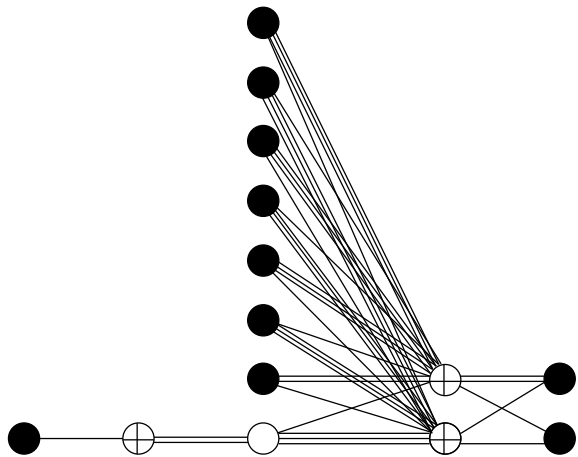}
\caption{Protograph of the rate-$4/5$ AR4JA code}
\label{fig:ar4ja}
\end{figure}

The rate-$4/5$ AR4JA protograph is shown in Figure~\ref{fig:ar4ja},
where dark circles are variable nodes, the white circle is the
punctured variable node, and plus circles are check nodes. The rate of the
code is $R=\frac{11-3}{11-1}=\frac{4}{5}$. The protograph can also be
represented by a protomatrix as follows:
\begin{equation}
H_{4/5}= \begin{pmatrix}
1& 0& 0& 0& 0& 0& 0& 0& 0& 0& 2\\
0 &1 &1& 1& 3& 1& 3& 1& 3& 1& 3\\
0& 1& 2& 2& 1& 3& 1& 3& 1& 3&1
\end{pmatrix}
\end{equation}
where non-zero entries indicate the number of edges connecting the
respective columns (variable nodes) and rows (check nodes). In this
protograph, the last column with the highest degree represents the
punctured node of the protograph. A derived LDPC code with the
information blocklength of $k=1024$ has the permutation size of
$128$. In other words, each variable or check node in the rate-$4/5$
protograph represents $128$ variable or check nodes with the same
degree in its derived LDPC code. To get a family of codes with rates
from $1/3$ to $4/5$, with each low-rate protograph, one variable and
one check node are added to the high-rate protograph in the rate level
above.  Thus, a family of rate-compatible codes can be built with
rates $R = \frac{8}{10+n}$ where $n=1,2,\ldots$ is the total number of
variable and check nodes adding to the daughter code.
%
%
A family of $15$ rate-compatible codes is designed
down to rate $1/3$. Equation~\eqref{eq:code13} shows the protomatrix
of the rate-$1/3$ code, which also contains the protomatrix of the
higher-rate codes as mentioned earlier.

\begin{figure}[h]
\begin{spacing}{1.2}
\setlength{\arraycolsep}{0.07cm}
\begin{align}
&H_{17\times 25} =\nonumber\\
&\begin{pmatrix}
1&0&0&0&0&0&0&0&0&0&2&0&0&0&0&0&0&0&0&0&0&0&0&0&0\\ 
0&1&1&1&3&1&3&1&3&1&3&0&0&0&0&0&0&0&0&0&0&0&0&0&0\\ 
0&1&2&2&1&3&1&3&1&3&1&0&0&0&0&0&0&0&0&0&0&0&0&0&0\\ 
0&0&0&0&0&0&1&1&1&1&1&1&0&0&0&0&0&0&0&0&0&0&0&0&0\\ 
0&0&0&0&0&1&1&1&1&1&2&0&1&0&0&0&0&0&0&0&0&0&0&0&0\\ 
0&0&0&0&1&1&1&1&1&1&2&0&0&1&0&0&0&0&0&0&0&0&0&0&0\\ 
0&0&1&0&0&0&1&1&1&1&2&0&0&0&1&0&0&0&0&0&0&0&0&0&0\\
0&0&1&0&0&1&1&0&1&0&2&0&0&0&0&1&0&0&0&0&0&0&0&0&0\\ 
0&0&1&0&0&0&1&0&1&0&1&0&0&0&0&0&1&0&0&0&0&0&0&0&0\\ 
0&1&0&0&0&0&1&0&1&1&1&0&0&0&0&0&0&1&0&0&0&0&0&0&0\\ 
0&0&0&0&0&1&0&0&1&1&2&0&0&0&0&0&0&0&1&0&0&0&0&0&0\\ 
0&0&0&0&1&0&1&0&1&0&1&0&0&0&0&0&0&0&0&1&0&0&0&0&0\\ 
0&0&0&0&1&1&0&0&1&0&2&0&0&0&0&0&0&0&0&0&1&0&0&0&0\\ 
0&1&0&0&0&1&0&0&1&0&1&0&0&0&0&0&0&0&0&0&0&1&0&0&0\\ 
0&0&1&0&0&0&1&0&0&0&1&0&0&0&0&0&0&0&0&0&0&0&1&0&0\\
0&1&0&0&0&0&1&0&1&0&1&0&0&0&0&0&0&0&0&0&0&0&0&1&0\\ 
0&1&1&0&1&0&0&0&0&0&1&0&0&0&0&0&0&0&0&0&0&0&0&0&1
\end{pmatrix}
\label{eq:code13}
\end{align}
\end{spacing}
\end{figure}

\begin{table}
\caption{Threshold ($E_b/N_0$ dB) of the new RC protograph codes}
\centering
\begin{tabular}{|c|c|c|c|}
\hline
Code & Protograph& Capacity & Gap to \\
Rate & threshold & threshold &capacity\\\hline
4/5  & 2.386&2.040&0.346 \\ \hline
8/11 &1.994&1.461&0.533  \\ \hline
2/3  & 1.554&1.059&0.495 \\ \hline
8/13 &1.166&0.760&0.406 \\ \hline
4/7  &0.919 &0.526&0.393 \\ \hline
8/15 &0.707&0.326&0.381 \\ \hline
1/2  &0.494&0.188&0.306 \\ \hline
8/17 &0.450&0.055&0.395 \\ \hline
4/9  &0.244&-0.064&0.308\\ \hline
8/19 &0.164&-0.150&0.314\\ \hline
4/5  &0.054 &-0.236&0.290\\ \hline
8/21 &0.021&-0.310&0.331\\ \hline
4/11 &-0.090&-0.390&0.300\\ \hline
8/23 &-0.148&-0.442&0.294\\ \hline
1/3  &-0.204&-0.507&0.303\\ \hline
\end{tabular}
\label{ta:new_proto}
\end{table}

In the search process, each new row of the protomatrix is constrained
to have a weight of at least four. This is because one graph edge from
the new check node must go to the new variable node, one or two to the
punctured node, and at least one more edge is necessary for connecting
to the remainder of the variable nodes.

In our search for good codes, we restrict to $\{0,1\}$ the number of
edges between new check nodes to old variable nodes, except for the
punctured variable node with the highest degree ($11^{th}$ column)
where the number of edges is restricted to $\{1,2\}$. This limitation
greatly simplifies the search space, and still yields good codes,
whose performance is described in the sequel.  For the search
technique we use a combination of constraints arising from threshold
considerations and error floor considerations~\cite{Thuy:TCOM10}, to
which is added a cost function component involving one point on the
FER curve at or around $10^{-5}$. More specifically, the search for
good codes uses the criteria in~\cite{Thuy:TCOM10}, but instead of
producing one optimum code according to those criteria, we now produce
a list containing $M$ codes with the $M$ best thresholds. Then from
among this set, the code with the steepest waterfall is chosen. Since
the code blocklength is relatively short, the required FER simulation
is fast and thus is not a serious burden in the search process. The
addition of this feature to the design is desirable because it is well
known that at short block lengths, the code threshold alone does not
adequately describe the code performance.

Our codes do not have the smallest threshold in their class; in fact
their threshold can be improved by providing more edges connecting to
the punctured variable node. However, such a modification will not
result in an improvement in the performance of short-length codes.
Allowing fewer edges connecting to the punctured variable node often
results in a superior performance at the $1$k block length although it
produces a higher threshold. As seen in Equation~\eqref{eq:code13}, in
our codes there are one or two edges connecting from new check nodes
to the punctured variable node.

Table~\ref{ta:new_proto} shows iterative decoding thresholds of the
new rate-compatible codes. The decoding threshold is computed by the
PEXIT algorithm~\cite{Liva2007}, showing that our codes have a
threshold gap between $0.3$-$0.53$ dB to capacity.


\section{Simulation Results}
\label{sec:numerical}

\begin{figure}
\centering
\includegraphics[width=3.5in]{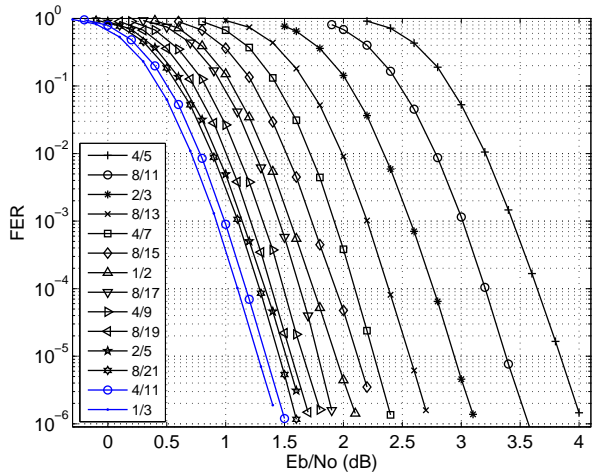}
\caption{FER performance of the new RC codes with information blocklength of $1$k.}
\label{fig:FER}
\end{figure}

 \begin{figure}
\centering
\includegraphics[width=3.5in]{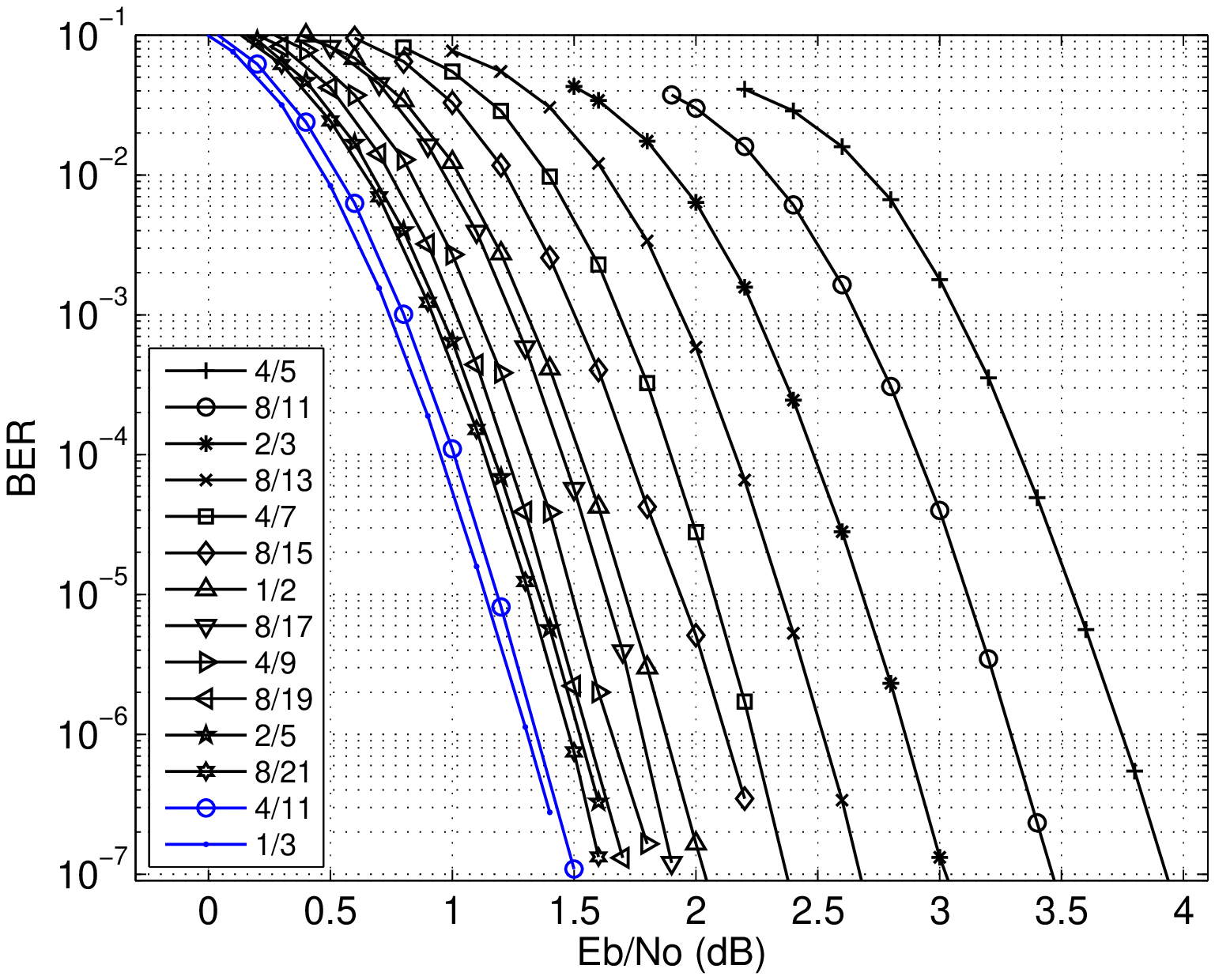}
\caption{BER performance of the new RC codes with information blocklength of $1$k.}
\label{fig:BER}
\end{figure}

\begin{figure}
\centering
\includegraphics[width=3.5in]{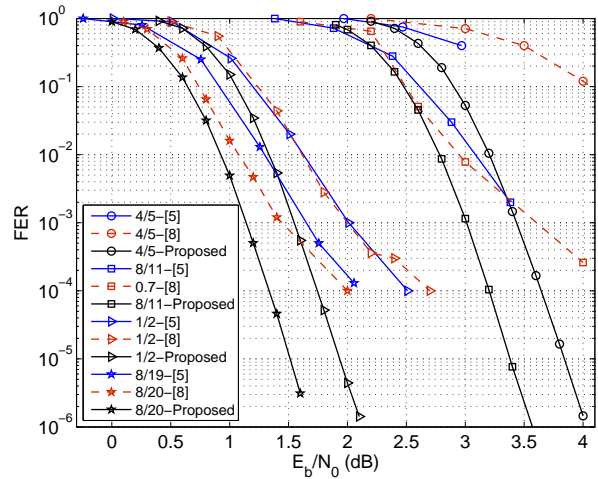}
\caption{FER performance comparison with codes reported
  in~\cite{Amir:TComL04,El-Khamy2009} with information blocklength of
  $1$k.}
\label{fig:compare}
\end{figure}


Our codes are derived from the protomatrix of
Equation~\eqref{eq:code13} in two lifting steps. First, the protograph
is lifted by a factor of $4$ using the progressive edge growth (PEG)
algorithm~\cite{Hu03_PEG} in order to remove all parallel edges. Then,
the second lifting using the PEG algorithm is performed to determine
a circulant permutation of each edge class that would yield the
desired code blocklength. In this paper, the size of circulant
permutation matrix is $32$, which yields an information block of $1024$
bits. The resulting LDPC codes have the girth of $6$.

Figure~\ref{fig:FER} and~\ref{fig:BER} show the FER and BER
performances of the proposed codes over the BI-AWGN channel. A
sum-product decoder with $8$ bit quantization is used. The maximum
number of iterations is set to $200$. No error floor is observed down
to FER$=10^{-6}$. The practical performance at FER$=10^{-5}$ shows a
gap of $1.7$ dB to capacity.

Figure~\ref{fig:compare} compares our code with codes reported
in~\cite{Amir:TComL04}.  We also compare with~\cite{El-Khamy2009} with
information blocklength of $1$k.~\cite{Amir:TComL04} designed random
rate-compatible codes by a hybrid (extension/puncturing) approach,
while~\cite{El-Khamy2009} designed rate-compatible protograph codes by
puncturing. To produce a clear plot, we only show the performance
curves of codes with rates $0.4$, $0.5$, $0.7$ and $0.8$. As seen in
the figure, previous codes exhibit error floors starting at
FER$=10^{-3}$, while the new code is free of error floors down to
FER=$10^{-6}$. For code rates below $0.7$, our codes outperform the
existing codes by $0.7$ dB at FER$=3\times 10^{-4}$. Our code at
rate-$0.72$ outperforms the rate-$0.7$ code
of~\cite{El-Khamy2009}. The rate-$0.8$ punctured codes
of~\cite{Amir:TComL04,El-Khamy2009} are easily outperformed by the new
codes. This highlights once again the known difficulties of designing
high-rate LDPC codes with puncturing at low- to moderate-blocklength
regime.

\section{Conclusion}
\label{sec:conclusion}
This paper presents a new family of rate-compatible structured LDPC
codes for short/moderate blocklengths. These codes have superior
performance compared with existing rate-compatible LDPC codes with
$1$k information blocklength in both waterfall and error floor
regions. 


\bibliographystyle{IEEEtran}
\bibliography{IEEEabrv,letter_refs} 
\end{document}